\documentclass[12pt]{article}
\usepackage[]{epsfig}
\begin{document}

\title{Network dependence of strong reciprocity}
\author{R. Vilela Mendes\thanks{%
Complexo Interdisciplinar, Universidade de Lisboa, Av. Gama Pinto 2,
1649-003 Lisboa, Portugal} \thanks{%
e-mail: vilela@cii.fc.ul.pt}}
\date{}
\maketitle

\begin{abstract}
Experimental evidence suggests that human decisions involve a mixture of
self-interest and internalized social norms which cannot be accounted for by
the Nash equilibrium behavior of Homo Oeconomicus. This led to the notion of
strong reciprocity (or altruistic punishment) to capture the human trait
leading an individual to punish norm violators at a cost to himself.

For a population with small autonomous groups with collective monitoring,
the interplay of intra- and intergroup dynamics shows this to be an adaptive
trait, although not fully invasive of a selfish population. However, the
absence of collective monitoring in a larger society changes the evolution
dynamics. Clustering seems to be the network parameter that controls
maintenance and evolution of the reciprocator trait.
\end{abstract}


\section{Homo oeconomicus versus homo reciprocans}

The assumption of self-interest as a motivation for social and economic
behavior is widely used as a guiding principle for social modeling. In a
game theory context the idea of maximization of self-interest leads to the
notion of (noncooperative) Nash equilibrium. A strategy is a Nash
equilibrium if no player can improve his payoff by changing his strategy,
when the strategies of the other players are fixed.

Given any environment situation, in a Nash equilibrium solution, each player
tries to maximize his gains regardless of what happens to the other players.
It is the rational expectations attitude of what has been called the \textit{%
Homo oeconomicus,} a notion which is at the basis of many theoretical
economics constructions. Whether this is a realistic notion when applied to
human societies is an important issue. Experiments have been carried out
and, in many cases, when played by human players, games have outcomes very
different from the Nash equilibrium points. An interesting case is the 
\textit{ultimatum game}\cite{Guth}. A simplified version of this game is the
following:

One of the players (the \textit{proposer }$P$) receives 100 coins which he
is told to divide into two non-zero parts, one for himself and the other for
the other player (the \textit{responder }$R$). If the responder accepts the
split $\left( R_{0}\right) $, it is implemented. If the responder refuses $%
\left( R_{1}\right) $, nothing is given to the players. Consider, for
example, a simple payoff matrix corresponding to two different proposer
offers $\left( P_{0}\textnormal{ and }P_{1}\right) $ 
\begin{equation}
\begin{tabular}{|c|c|c|}
\hline
& $R_{0}$ & $R_{1}$ \\ \hline
$P_{0}$ & $a,c$ & $0,0$ \\ \hline
$P_{1}$ & $b,b$ & $0,0$ \\ \hline
\end{tabular}
\label{1.1}
\end{equation}
with $a\gg c$, $a+c=2b$ (for example $a=99,c=1,b=50$).

The unique Nash equilibrium is $\left( P_{0},R_{0}\right) $, corresponding
to the payoffs $\left( a,c\right) $. However, when the game is played with
human players, such greedy proposals are most often refused, even in
one-shot games where the responder has no material or strategic advantage in
refusing the offer. Based on this and similar results in other situations
(public goods games, etc), Bowles and Gintis\cite{Bowles1} \cite{Bowles2}
developed the notion of strong reciprocity (\textit{Homo reciprocans}\cite
{Bowles4}) as a better model for human behavior. \textit{Homo reciprocans
would come to social situations with a propensity to cooperate and share but
would respond to selfish behavior on the part of others by retaliating, even
at a cost to himself and even when he could not expect any future personal
gains from such actions}. This should be distinguished from cooperation in a
repeated game or reciprocal altruism or other forms of mutually beneficial
cooperation that can be accounted for in terms of self-interest.

The same authors, in collaboration with a group of anthropologists,
conducted a very interesting ``ultimatum game experiment'' in many
small-scale societies around the world\cite{Bowles3}. Homo oeconomicus is
rejected in all cases and consistently different results are obtained in
different societies, the players' behavior being strongly correlated with
existing social norms and the market structure in their societies. This and
other experiments \cite{Gintis2} \cite{Gintis4} strongly suggest that human
decision problems involve a mixture of self-interest and a background of
(internalized) social norms \cite{Gintis3} \cite{Kaplan}.

Strong reciprocity is a form of altruism \cite{Fehr} in that it benefits
others at the expense of the individual that exhibits this trait. Monitoring
and punishing selfish agents or norm violators is a costly (and dangerous)
activity without immediate direct benefit to the agent that performs it. It
would be much better to let others do it and to reap the social benefits
without the costs.

Strong reciprocator agents contribute more to the group than selfish ones
and they sustain the cost of monitoring and punishing free-riders. For this
reason it was thought that the strong reciprocity trait could not invade a
population of self-interested agents, nor could it be maintained in a stable
population equilibrium. To counter this belief, Bowles and Gintis \cite
{Bowles2} developed a simple (mean-field type) model that might apply to the
structure of the small hunter-gatherer bands of the late Pleistocene. Taking
the view that the \textit{strong reciprocity} trait has a genetic basis,
this would be a period long enough to account for a significant development
in the modern human gene distribution. The model would give an evolutionary
explanation of the phenomenon. Of course, if instead of gene-based, strong
reciprocity is culturally inherited, emergence and (or) modification of this
trait could be much faster.

Because I intend to explore the influence of the social (network) structure
on the evolution of strong reciprocity, I will start by discussing a
simplified version of the Bowles-Gintis model. The main simplification is
that migration in and out of the evolving group to an outside pool of agents
is not considered. The consideration of these migrations may be of interest
for a realistic picture of the hunter-gatherer bands of the Pleistocene, but
not for the general picture of strong reciprocity in a wider society. By
simplifying and somewhat enlarging the punishment scenario (beyond
ostracism) of the Bowles-Gintis model and framing it as a replicator
one-dimensional map, a clear view is obtained of its dynamical aspects.

\section{Emergence of strong reciprocity. The Bowles-Gintis model}

One considers a population of size $N$ with two species of agents, one
denoted \textit{reciprocators} (R-agents) and the other \textit{%
self-interested} (S-agents). In a \textit{public goods} activity each agent
can produce a maximum amount of goods $q$ at cost $b$ (with goods and costs
in fitness units). The benefit that an S-agent takes from shirking public
goods work is the cost of effort $b\left( \sigma \right) $, $\sigma $ being
the fraction of time the agent shirks. The following conditions hold 
\begin{equation}
b\left( 0\right) =b,\qquad b\left( 1\right) =0,\qquad b^{^{\prime }}\left(
\sigma \right) <0,\qquad b^{^{\prime \prime }}\left( \sigma \right) >0
\label{2.1}
\end{equation}
Furthermore $q\left( 1-\sigma \right) >b\left( \sigma \right) $ so that, at
every level of effort, working helps the group more than it hurts the worker.

For $b\left( \sigma \right) $ one chooses 
\begin{equation}
b\left( \sigma \right) =\frac{2}{2\sigma -1+\sqrt{1+4/b}}-\frac{2}{1+\sqrt{%
1+4/b}}  \label{2.2}
\end{equation}
which satisfies the constraints (\ref{2.1}).

R-agents never shirk and punish each free-rider at cost $c\sigma $, the cost
being shared by all R-agents. For an S-agent the estimated cost of being
punished is $s\sigma $, punishment being ostracism or some other fitness
decreasing measure. $s$ is the weight given by an S-agent to the punishment
probability. It may or may not be the same as the actual fitness cost of
punishment. Each S-agent chooses $\sigma $ (the shirking time fraction) to
minimize the function 
\begin{equation}
B\left( \sigma \right) =b\left( \sigma \right) +sf\sigma -q\left( 1-\sigma
\right) \frac{1}{N}  \label{2.3}
\end{equation}
$f$ being the fraction of R-agents in the population, $f\sigma $ is the
probability of being monitored and punished. The last term is the agent's
share of his own production. The value $\sigma _{S}$ that minimizes $B\left(
\sigma \right) $ is 
\begin{equation}
\sigma _{S}=\max \left( \min \left( \frac{1}{2}-\sqrt{\frac{1}{4}+\frac{1}{b}%
}+\frac{1}{\sqrt{sf+\frac{q}{N}}},1\right) ,0\right)  \label{2.4}
\end{equation}

The contribution of each species to the population in the next time period
is proportional to its fitness given by 
\begin{equation}
\begin{array}{lll}
\pi _{S}^{^{\prime }}\left( f\right) & = & q\left( 1-\left( 1-f\right)
\sigma _{S}\right) -b\left( \sigma _{S}\right) -\gamma f\sigma _{S} \\ 
\pi _{R}^{^{\prime }}\left( f\right) & = & q\left( 1-\left( 1-f\right)
\sigma _{S}\right) -b-c\left( 1-f\right) \frac{N\sigma _{S}}{Nf}
\end{array}
\label{2.5}
\end{equation}
for S- and R-agents. The baseline fitness is zero, that is, $\pi _{S,R}=\max
\left( \pi _{S,R}^{^{\prime }},0\right) $.

The first term in both $\pi _{S}^{^{\prime }}$ and $\pi _{R}^{^{\prime }}$
is the benefit arising from the produced public goods and the second term
the work effort. The last terms represent the fitness cost of punishment for
S-agents and the cost incurred by R-agents.

$\gamma =1$ corresponds to ostracism from the group, other values to general
coercive measures affecting the fitness of S-agents. The last term in $\pi
_{R}^{^{\prime }}$ emphasizes the collective nature of the punishment.
Notice however the improbable heavy punishing burden put on reciprocators
when in small number.

Finally one obtains\footnote{%
Here replicator dynamics is used for the population evolution. Notice that
Bowles and Gintis\cite{Bowles2} use a different (incremental) dynamics.} a
one-dimensional map for the evolution of the fraction of R-agents 
\begin{equation}
f_{\textnormal{new}}=f\frac{\Pi _{R}\left( f\right) }{\left( 1-f\right) \Pi
_{S}+f\Pi _{R}\left( f\right) }  \label{2.6}
\end{equation}

\begin{figure}[htb]
\begin{center}
\psfig{figure=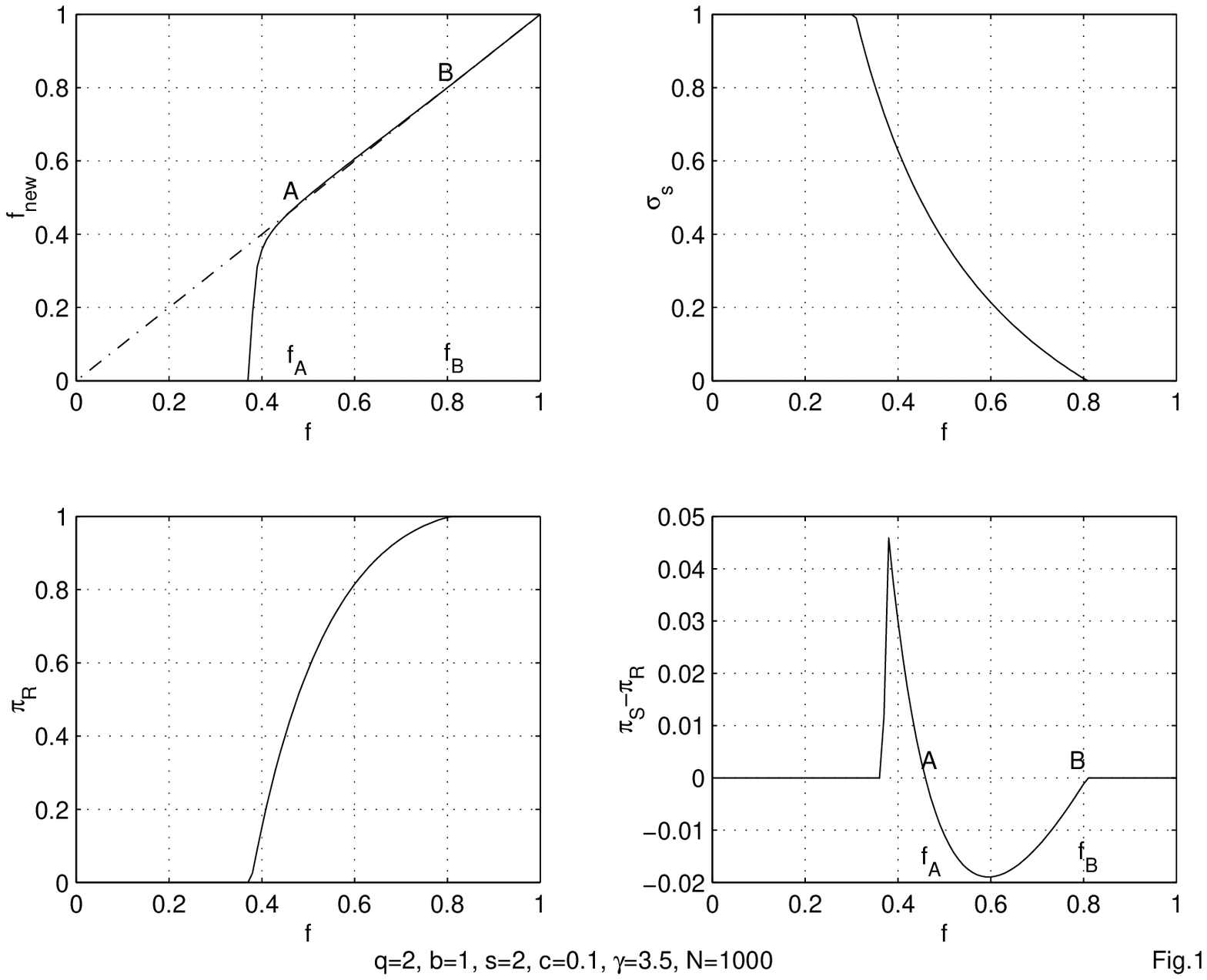,width=8truecm}
\end{center}
\caption{}
\end{figure}

\begin{figure}[htb]
\begin{center}
\psfig{figure=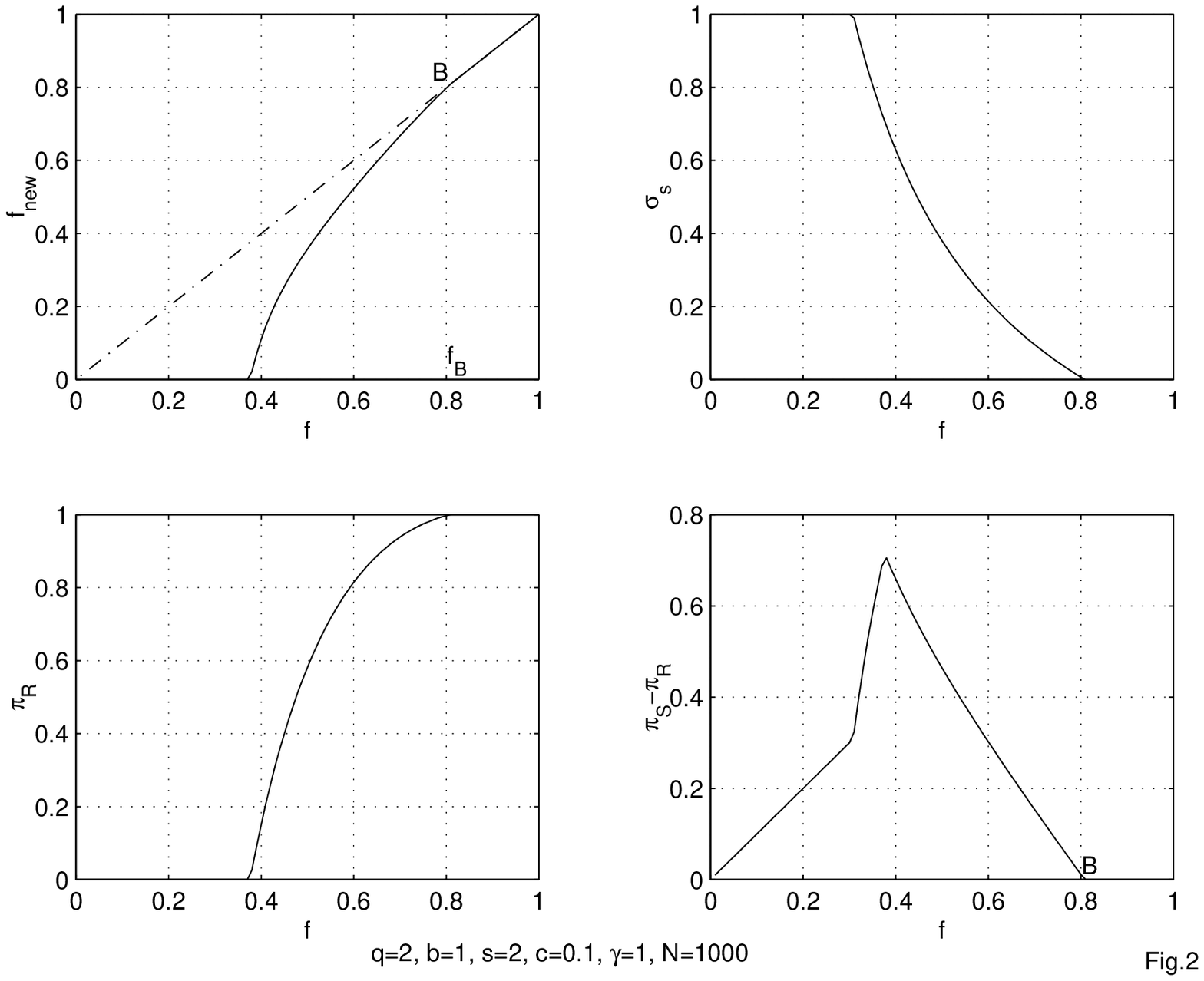,width=8truecm}
\end{center}
\caption{}
\end{figure}

Figs. 1 and 2 display this map, as well as $\sigma _{S}\left( f\right) $, $%
\Pi _{R}\left( f\right) $ and $\Pi _{S}\left( f\right) -\Pi _{R}\left(
f\right) $ for two different values of $\gamma $, the other parameters being
the same. They show the general behavior of the map in Eq.(\ref{2.6}). If $%
\gamma $ (the fitness impact of punishment) is large enough, the map has an
unstable fixed point $A$ at $f_{A}$ and a left-stable one $B$ at $f_{B}$.
Between $f_{B}$ and $1$ there is a continuum of marginally stable fixed
points. For smaller $\gamma $ the region between $f_{A}$ and $f_{B}$ (where $%
\Pi _{S}-\Pi _{R}$ is negative) disappears and only the marginally stable
fixed points remain. In both cases the asymptotic behavior corresponds
either to $f=0$ (and $\sigma _{S}=1$) or to $f$ between $0$ and $1$ but $%
\sigma _{S}=0$. That is, in this second case, both reciprocators and
shirkers remain in the population but shirkers choose not to shirk because
the minimum of $B\left( \sigma \right) $ is at $\sigma _{S}=0$.

For an initial $f$ smaller than $f_{A}$ the fraction of reciprocators falls
very rapidly to zero. This reflects the (maybe unrealistic) fact that in
this case a very small number of reciprocators has to carry the burden of
punishing very many shirkers.

Hence, from the point of view of intragroup dynamics, either reciprocators
are completely eliminated from the population or they remain in equilibrium
with a probably large number of shirkers, which do not shirk for fear of
being punished. Therefore intragroup dynamics, by itself, cannot explain how
the reciprocator attitude might have become a dominant trait. However when
very many groups are considered, for example assembled at random from a pool
containing both reciprocators and shirkers \cite{Sethi} \cite{Sethi2}, then
only the groups that contain at the start a fraction $f$ greater than $f_{A}$
will have in the end a nonzero fitness. In all others, S-agents invade the
population and suffer a ``tragedy of the commons'' situation with final zero
fitness. Therefore from an intergroup dynamics perspective the groups with
reciprocators tend to dominate and impose an above average predominance of
the reciprocator trait.

Although the model, together with intergroup dynamics, explains why strong
reciprocity is an adaptive trait \cite{Fehr2}, the marginally stable nature
of the (above $f_{B}$) fixed points also suggest that the shirker trait is
never eliminated and will remain in the population.

Small independent groups assembling and disassembling is a likely scenario
for the development of the reciprocator trait. In this sense the
hunter-gatherer bands of the Pleistocene might have indeed provided the
appropriate environment for the evolution of the trait, whether gene-coded
or culturally-inherited.

It is well known that group size affects monitoring in public goods
provision \cite{Carpenter}. Therefore, a natural question is what happens
when, later on, the Pleistocene reciprocators and their fellow shirkers
become imbedded into a larger society. Monitoring and punishment of shirkers
by reciprocators necessarily looses its global collective nature. Once
monitoring looses its global nature, it becomes the business of the
neighbors of the shirker. In addition to the individual cost of monitoring
and (or) punishing free-riders, such punishing requires an amount of force
that, in particular, insures the effectiveness of the punishment and on the
other hand puts the punisher safe from direct retaliation from the violator.
This is one of the reasons for the creation of central authorities for this
purpose. However if central authorities have enough force to implement
punishment without retaliation, they are at times quite ineffective at
monitoring. Also laws and central authorities, on the role of reciprocators
play, a role in the control of serious offenses, but not so much on the day
to day monitoring of public goods work. Therefore in a large society the
nature of the control performed by the neighbors is certainly going to play
a role on the evolution of the reciprocator trait.

If the trait is genetically encoded, maybe the wider societies developed by
modern man had no time to make significant changes on its structure. However
if it is (at least in part) culturally inherited then a much shorter time
scale may be involved. What about the big city tales of a guy being mugged
in full daylight while a crowd of passersby moves along quite indifferent to
the event? Is it the ($1-f$) remnants of non-reciprocators in the population
or are we watching the emergence of Homo Oeconomicus in his full glory? Or
is it something else?

\section{Network dependence of strong reciprocity}

To explore the possible effect of the social network structure on the
evolution of strong reciprocity I will consider a agent-based model, which
later on will be interpreted in a mean field sense similar to the model in
Section 2.

As before one considers R-agents and S-agents and the monitoring function
performed by R-agents is kept at the neighbors level. However punishment is
only implemented if at least two neighbors are willing to do so. It is the
same as to say that punishing a norm-violator cannot be an individual
action, but requires a minimal social power and consensus. The need to be
close to monitor and the need for agreement of at least two neighboring
reciprocators to implement punishment, immediately suggests that the
structure of the network is going to play a role on the evolution of the
group. The following is the mathematical coding of this idea:

As before one has two agent species (S-agents and R-agents), the fraction of
R-agents being $f$. The agents are placed in a network where, on average,
each agent is connected to $k$ other agents. $k$ is called the \textit{degree%
} of the network. To the whole population of dimension $N$ one associates 3 $%
N-$dimensional vectors, $Wk$, $Pu$, $Cpu$. $Wk$ is called the\textit{\ work
vector,} $Pu$ the \textit{punishment vector} and $Cpu$ the \textit{cost of
punishment vector}.

The link structure of the network is chosen as in the $\beta -$model of
Watts and Strogatz\cite{Watts1} \cite{Watts2}. Starting from a regular ring
structure where each node is symmetrically connected to its $k$ closest
neighbors, each link is examined in turn and, with probability $\beta $,
replaced by a random link to some other node in the network.

At time zero, $fN$ R-agents and $\left( 1-f\right) N$ S-agents are placed at
random in the network. The local neighborhood of agent $i$, that is the set
of other agents connected to $i$, is denoted $\Gamma _{i}$. The entries of
the vectors $Wk$, $Pu$, $Cpu$ are then computed as follows:

\# For the $Wk$ vector

R-agents; $Wk\left( i\right) =1$

S-agents; $Wk\left( i\right) =\frac{n_{R}\left( i\right) }{k}$, where $%
n_{R}\left( i\right) $ is the number of R-agents connected to this S-agent, $%
n_{R}\left( i\right) =\#\left\{ j:j\in R,j\in \Gamma _{i}\right\} $

\# For the $Pu$ vector

R-agents; $Pu\left( i\right) =0$

S-agents;\ $Pu\left( i\right) =n_{P}\left( i\right) \left( 1-Wk\left(
i\right) \right) $, where $n_{P}\left( i\right) $ is the number of pairs of
R-agents in $\Gamma _{i}$ which are also neighbors among themselves, $%
n_{P}\left( i\right) =\#\left\{ \left( j,k\right) :\left( j,k\right) \in
R,\left( j,k\right) \in \Gamma _{i},j\in \Gamma _{k}\right\} $ and $\left(
1-Wk\left( i\right) \right) $ is the shirking fraction.

\# For the $Cpu$ vector

R-agents; $Cpu\left( i\right) =\sum_{k\in S}n_{C}\left( i,k\right) \left(
1-Wk\left( k\right) \right) $ where $n_{C}\left( i,k\right) $ is the number
of times that the agent $i$ is in a R-pair punishing an S-agent $k$, $%
n_{C}\left( i,k\right) =\#\left\{ \left( i,j\right) :k\in S,\left(
i,j\right) \in R,\left( i,j\right) \in \Gamma _{k},j\in \Gamma _{i}\right\} $

S-agents; $Cpu\left( i\right) =0$

Summarizing:

Each reciprocator, on detecting an S-agent $k$, looks for another
reciprocator in his own neighborhood also connected to $k$. If he finds one
he punishes $k$ by an amount proportional to the fraction of shirking. An
S-agent may be punished several times by all different pairs of
reciprocators in his neighborhood.

The amount of work that an S-agent does is inversely proportional to the
number of reciprocators in his neighborhood. However lack of communication
between neighboring reciprocators may make the probability of punishment
much smaller.

The (average) fitness of R-agents and S-agents is 
\begin{equation}
\pi _{R}^{^{\prime }}=\frac{q}{N}\sum_{all}Wk\left( i\right) -\frac{b}{fN}%
\sum_{i\in R}Wk\left( i\right) -\frac{c}{fN}\sum_{i\in R}Cpu\left( i\right) 
\label{3.1}
\end{equation}
\begin{equation}
\pi _{S}^{^{\prime }}=\frac{q}{N}\sum_{all}Wk\left( i\right) -\frac{b}{%
\left( 1-f\right) N}\sum_{i\in S}Wk\left( i\right) -\frac{\gamma }{\left(
1-f\right) N}\sum_{i\in S}Pu\left( i\right)   \label{3.2}
\end{equation}
The baseline fitness is zero, that is 
\begin{equation}
\pi _{R,S}=\max \left( \pi _{R,S}^{^{\prime }},0\right)   \label{3.3}
\end{equation}
Once the fitness is computed the replicator equation 
\begin{equation}
f_{\textnormal{new}}=f\frac{\pi _{R}}{f\pi _{R}+\left( 1-f\right) \pi _{S}}
\label{3.4}
\end{equation}
is applied and a new cycle starts with a new random distribution, on the
network, of $Nf_{\textnormal{new}}$ R-agents and $N\left( 1-f_{\textnormal{new}}\right) $
S-agents.

\begin{figure}[htb]
\begin{center}
\psfig{figure=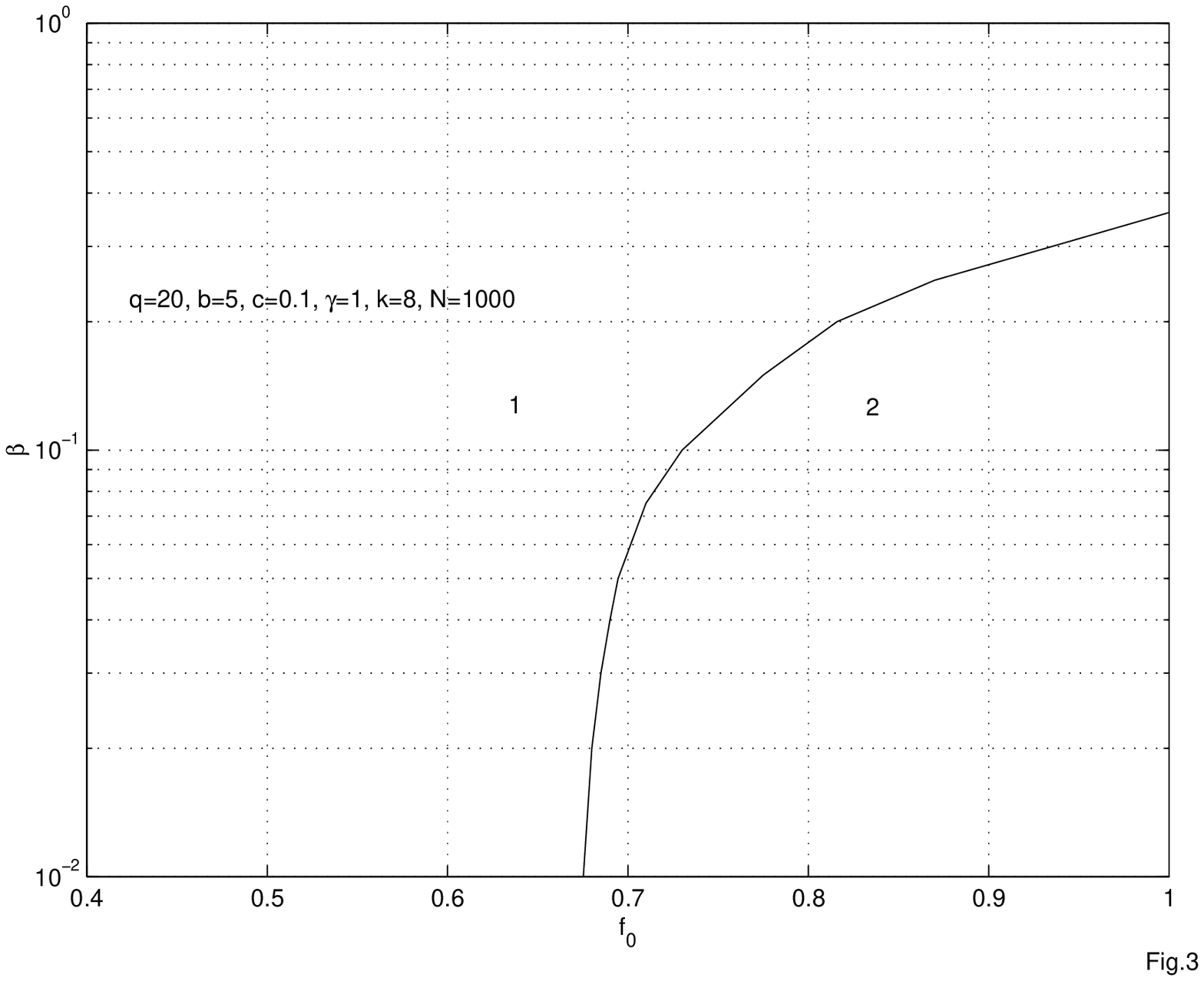,width=8truecm}
\end{center}
\caption{}
\end{figure}

Running this agent model for several values of $\beta $ and, in each case,
for random initial $f_{0}$'s one finds two separate regions in the $\left(
f_{0},\beta \right) $ plane (Fig.3). In region 1 the evolution drives $f$
towards zero as well as the overall fitness $\pi $ (Example in Fig.4a)

\begin{figure}[htb]
\begin{center}
\psfig{figure=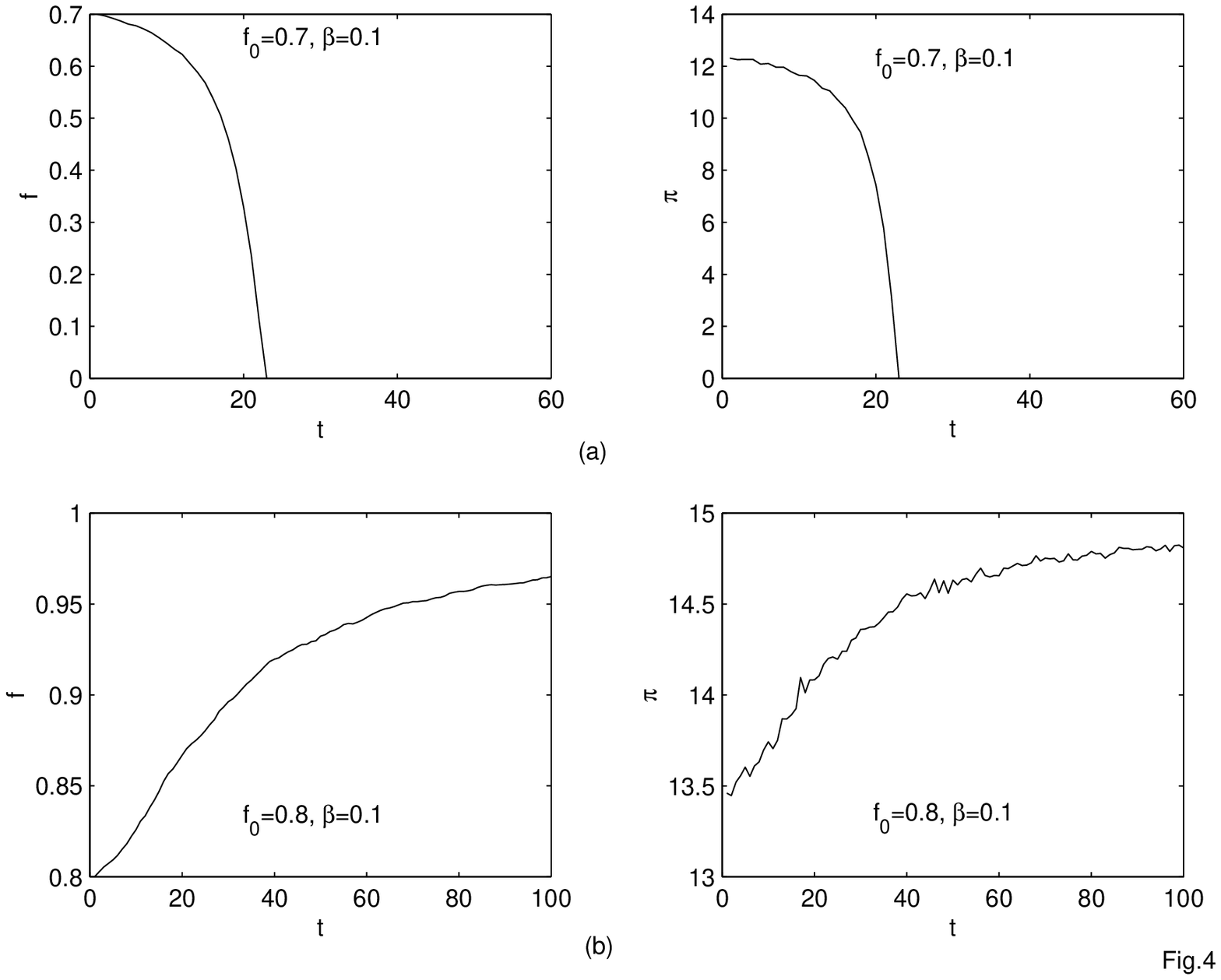,width=8truecm}
\end{center}
\caption{}
\end{figure}

\begin{equation}
\pi =f\pi _{R}+\left( 1-f\right) \pi _{S}  \label{3.5}
\end{equation}
In region 2 there is an asymptotic nonzero value for $f$ and for the fitness
(Example in Fig.4b).

As $\beta $ increases it becomes less likely to have a stable nonzero $f$.
the origin of this effect is clear. Although $\beta -$rewiring maintains the
average degree of the network, the probability of two neighbors of an agent
to be themselves neighbors decreases. Therefore it becomes increasingly
difficult for reciprocators to find local consensus for the punishment of
S-agents.

The average probability of two R-neighbors of a network node in S to be
themselves neighbors, is called the (relative) \textit{clustering coefficient%
}, 
\begin{equation}
C_{R}=\overline{\frac{n_{P}\left( i\right) }{\left( 
\begin{array}{c}
\#\left\{ \Gamma _{i}\cap R\right\} \\ 
2
\end{array}
\right) }}  \label{3.6}
\end{equation}
$\left( 
\begin{array}{c}
\#\left\{ \Gamma _{i}\cap R\right\} \\ 
2
\end{array}
\right) $ being the maximum possible number of links between the R-neighbors
of S-agent $i$. The network clustering coefficient is related to the notion
of transitivity used in the sociological literature.

For the $\beta -$rewiring model, the clustering may be estimated from the
number $\Phi $ of shortcuts which in this case is proportional to $\beta $ 
\cite{Watts2}. 
\begin{equation}
C_{\beta }\left( \Phi ,k\right) =\frac{\frac{3}{4}\left( 1-\Phi \right)
^{2}\left( k-\frac{2}{3}\right) -\left( 1-\Phi \right) }{k-1}  \label{3.7}
\end{equation}
Therefore a mean field version of the agent model may be written as follows 
\begin{equation}
\Pi _{R}^{^{\prime }}=q\left( 1-\left( 1-f\right) \sigma _{S}\left( f\right)
\right) -b\left( \sigma _{S}\left( f\right) \right) -\gamma fC_{\beta
}\left( \Phi ,fk\right) \sigma _{S}\left( f\right)  \label{3.8}
\end{equation}
\begin{equation}
\Pi _{S}^{^{\prime }}=q\left( 1-\left( 1-f\right) \sigma _{S}\left( f\right)
\right) -b-c\left( 1-f\right) \frac{fk}{2}C_{\beta }\left( \Phi ,fk\right)
\sigma _{S}\left( f\right)  \label{3.9}
\end{equation}
Notice the term $fk$ in $C_{\beta }\left( \Phi ,fk\right) $ and in the cost
of punishment term in $\Pi _{S}^{^{\prime }}$. It reflects the fact that
neighborhood relations for reciprocators are to computed on their subnetwork
of size $fN$.

$b\left( \sigma \right) $ is as in Eq.(\ref{2.2}) with $\sigma _{S}$ being
computed to minimize 
\begin{equation}
B\left( \sigma \right) =b\left( \sigma \right) +sfC_{\beta }\left( \Phi
,fk\right) \sigma -q\left( 1-\sigma \right) \frac{1}{N}  \label{3.10}
\end{equation}

This mean field version gives results identical to the agent-based model.
Clustering appears therefore as the determining network parameter driving
the evolution of the reciprocator trait.

\section{Conclusions}

1 - With a structure of small groups with collective monitoring of the
agents' activities, the fitness difference between groups with a sizable
amount of reciprocators and those where they have disappeared, makes the
emergence of the strong reciprocity trait a likely event.

However rather than being completely invaded by reciprocators, maintenance
of a certain amount of self-interested types is also likely, which only
cooperate for fear of being punished. If, at a later stage, the social
structure changes, they may be a source of instability and invade the
population.

2 - In a large population, monitoring of the public goods behavior of the
agents cannot be a fully collective activity, rather being the chore of
those in close contact with the free-riders. Punishment of free-riders also
requires a certain amount of local consensus among reciprocators. Therefore
the clustering nature of the society may play an important role in the
maintenance and evolution of the reciprocator trait.

Maybe the indifferent passersby that let the poor guy be mugged are not yet
homo oeconomicus. Maybe they are just reciprocators in the middle of
strangers with whom they do not communicate nor trust. A clustering problem.

3 - Culturally-inherited traits may have a much faster dynamics than
gene-based ones. Modern societies are ``small worlds'' in the sense of short
path lengths but not necessarily in the sense of also maintaining a high
degree of clustering. Therefore if the reciprocator trait has a high
cultural component, it may very well happen that, eventually, we will see
homo oeconomicus leaving the benches of economy classes for a life on the
streets.

\end{document}